\documentclass[runningheads,a4paper]{llncs}

\usepackage{url}
\usepackage{amsmath}
\usepackage{amssymb}
\usepackage{multirow}
\usepackage{graphicx}
\usepackage{graphics}
\usepackage{epstopdf}
\usepackage{color}
\usepackage{tabularx}

 \usepackage{floatrow}
 \usepackage{blindtext}
 \usepackage{wrapfig}
  \usepackage{subfig}
 \usepackage{afterpage}

\def\etal{{\it et al.}}
 
 \usepackage{url}
 \urldef{\mailsa}\path|{dfay.bournemouth.ac.uk}|
  
\newcommand{\keywords}[1]{\par\addvspace\baselineskip\noindent\keywordname\enspace\ignorespaces#1}

\begin{document}

\mainmatter
\title{An exploration of fetish social networks and communities.}

\titlerunning{An exploration of fetish social networks and communities.}

\author{Damien Fay%
\and Hamed Haddadi \and Michael C. Seto\\
Han Wang \and Christoph Carl Kling}
\authorrunning{Fay, Haddadi, Seto, Wang, Kling}

\institute{Department of Computing, Bournemouth, UK\\
School of Elec. Eng. and Comp. Sci., Queen Mary University, UK\\
The Royal's Institute of Mental Health Research, Ottawa, Canada\\
National University of Ireland, Galway\\
GESIS -- Leibniz Institute for the Social Sciences, Cologne, Germany\\
\mailsa { }
}
\toctitle{Lecture Notes in Computer Science}
\maketitle

%
%

\begin{abstract}
Online Social Networks (OSNs) provide a venue for virtual interactions and relationships between individuals. In some communities, OSNs also facilitate arranging offline meetings and relationships. FetLife, the world’s largest anonymous social network for the BDSM, fetish and kink communities, provides a unique example of an OSN that serves as an interaction space, community organizing tool, and sexual market. In this paper, we present a first look at the characteristics of European members of Fetlife, comprising 504,416 individual nodes with 1,912,196 connections. We looked at user characteristics in terms of gender, sexual orientation, and preferred role. We further examined the topological and structural properties of groups, as well as the type of interactions and relations between their members. Our results suggest there are important differences between the FetLife community and conventional OSNs. The network can be characterised by complex gender based interactions both from a sexual market and platonic viewpoint which point to a truly fascinating social network.



\end{abstract}

%
%

\section{Introduction}
\label{sec:intro}

Social interaction is motivated at the individual level in need for power, prestige and approval \cite{turner1988theory} which are expressed in modern life in activities such as business, friendship/emotional learning exchange, and knowledge exchange; and from an evolutionary perspective the need to seek a mate. This latter function of a social network is known as the \emph{sexual market} and every social network has a secondary function as a sexual market, although disaggregating this function from others can be challenging~\cite{laumann2004sexual}. In the last decade, Online Social Networks (OSNs) have become a focal point of the web and the most popular activity of individuals online. There are a large number of popular OSNs and a large body of research focuses on a variety of OSNs. Despite a large number of papers on analysis of large scale OSNs~\cite{Mislove,cha2012world}, and a large number of social science papers on social relationships, sexuality and orientations~\cite{journals/socnet/MollenhorstVF08}\cite{Joyal}, there have not been any academic papers which have examined online social networks focused on variations in sexual orientations and interests. 

In this paper, we take a first look at the anonymised profiles of the European users of the most popular fetish website, and ask if the characteristics of the network are different from those of a conventional OSN. This is a rich dataset of over half a million users and captures patterns of traditionally secret interests and behaviours. We do so by comparing the topological characteristics with those of popular social networks as reported by Mislove \etal~\cite{Mislove}. We choose this online fetish network as it is oriented towards friendships, social groups, and arranging events, where the social is primary and sexual market is secondary but explicitly included (unlike, say, Facebook or other non-dating OSNs). It is important to social scientists and psychologists to understand whether a social network is also present or not required. As FetLife reveals sexuality in a social context it allows us to understand sexual networks in a way that dating sites such as Tinder, Grindr etc might not allow; this is also vital for creating models for the spread of sexually transmitted infections~\cite{Niekamp2013223}.

We use our large dataset to assess the properties of these multi-relationship networks, where a user can have a number of different types of relationships with others.\footnote{In the interest of space and scientific focus, we encourage the readers to see~\cite{averagejoe,kinkytom} for a description of different types of fetish relationships.} We base our analysis of the structure of the graph on work by Laumann~\etal~\cite{laumann2004sexual}, who use self reports and  assess individuals' roles and economic factors in sex markets, using four neighbourhoods in Chicago and highlight the role of brokers and third parties in this exchange. Our dataset uses the largest broker out there, the world's most popular fetish site, as a benchmark for analysis of the online version of this market. Understanding the nature of the interactions is also important for real and cyber crime investigations, as the privacy and safety of users could also be compromised by malicious users of such websites.\footnote{\url{http://sexandthe405.com/fetlife-is-not-safe-for-users/}}

\section{Online Fetish Networks}
\label{sec:fetlife}

We collected our data from FetLife,\footnote{\url{https://fetlife.com/}} the most popular Social Network for the BDSM, Fetish, and kink communities, with millions of users worldwide. The fetish community has grown rapidly in recent years and now consists of a diverse collection of people whose interests cover a broad spectrum including, fashion, burlesque, a nightclub scene, particular types of music and of course a focus on sexual experimentation. As in Facebook, the interaction of the community is both real-world and virtual with a large collection of real-world events attended by members; contrary to expectations, FetLife it is not a paid dating website. For example, there is no ``search" functionality within the website for specific types of members, e.g., based on interests, or over user information fields (height, weight, age, location, fetish commonalities, other personal  information). However the site is used as a bootstrapping mechanism for social events, workshops, and parties which are organised regionally. Members create a personal profile, similar to most OSNs, specify their gender, age, role, orientation, and list the fetishes they are interested in or are curious about. The users are organised into tens of thousands of groups, and thousands of events are arranged annually through the website. Users pay particular attention to the experience of the group members and event organisers and hence these individuals play a central role in the community. In essence, FetLife is a niche OSN. BDSM is a sexual interest or subculture attractive to a minority~\cite{Richters2008}. What makes FetLife unique particularly interesting for OSN analysts is that this website observes sexual interaction (present in dating websites, absent in typical social networks such as Facebook) but in the presence of a social context (absent in dating websites).

\section{Data collection}
\label{sec:data}

We collected our data from the European members of FetLife during the early months of 2014. The data includes anonymized (at the time of collection) user IDs, relationship types, and number of friends. In order to comply with the website policy and ethics approval requirements, we did not crawl any names, details of friends, pictures, posts, or other personally identifiable information available on the site. Since it is mandatory for users to be a member of a single geographic area (usually county/borough level depending on the population density), our crawler used the location area codes of the website as its seed and we collected the mentioned details about every single individual in the European section of the website.


Overall, there are 504,416 individual nodes in our dataset, with 1,912,196 connections. The main connected component is comprised of just over 156K nodes, and the rest of the users are mainly isolated or small groups of maximum size 20. At the time of collection, there were 35,153 groups in the dataset, with just over 26k single nodes. Although this is a sample of the population and only captures the individuals who chose to be on a fetish OSN, this data is more inclusive and less biased than the offline club members or those who self-identify for sample surveys in existing literature~\cite{dawson2014paraphilic,Richters2008}. The perceived anonymity online and low (essentially zero) cost of entry into Fetlife means more individuals might be active online than joining actual clubs, going to local BDSM themed parties or self-identifying to researchers at universities.


\section{Demographic analysis}
\label{subsec:demo_anal}

\begin{figure}[t!]
 \centering
 \includegraphics[width=0.75\textwidth]{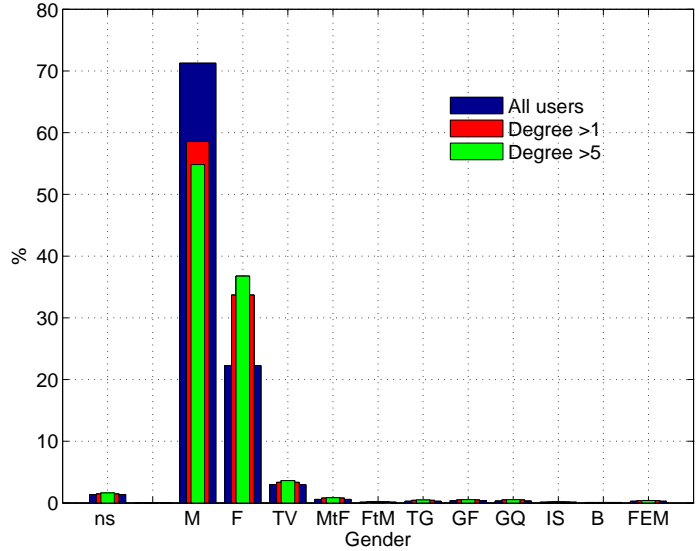} 
 \caption{Distribution of genders for all users, users with $>1$ friends, and $>5$ friends.}\label{f:gender_per}
 \end{figure}
\begin{figure}[t!]
 \centering
 \includegraphics[width=0.75\textwidth]{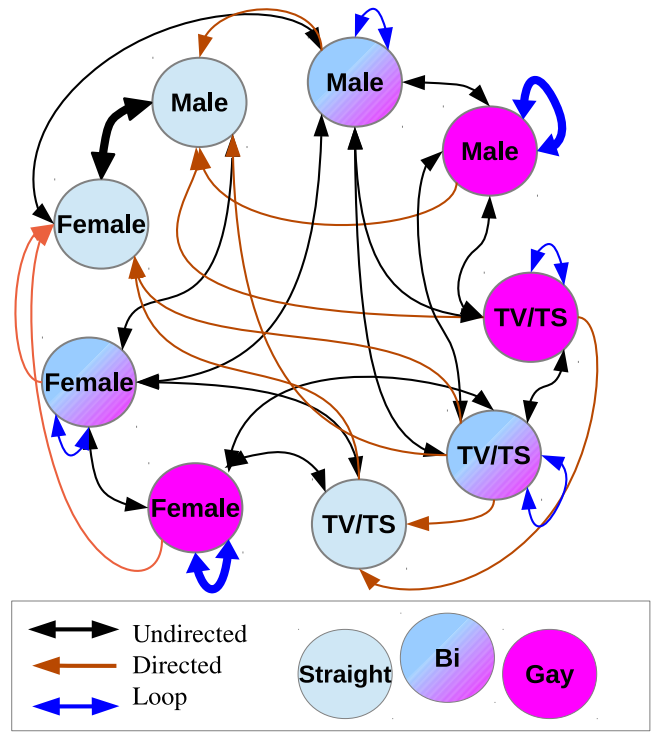} 
 \caption{Graph of potential partners. Note that some links are directed, and the graph is not complete.}\label{f:GenderDrawing}
 \end{figure}
\begin{figure}[t!]
 \centering
 \includegraphics[width=0.75\textwidth]{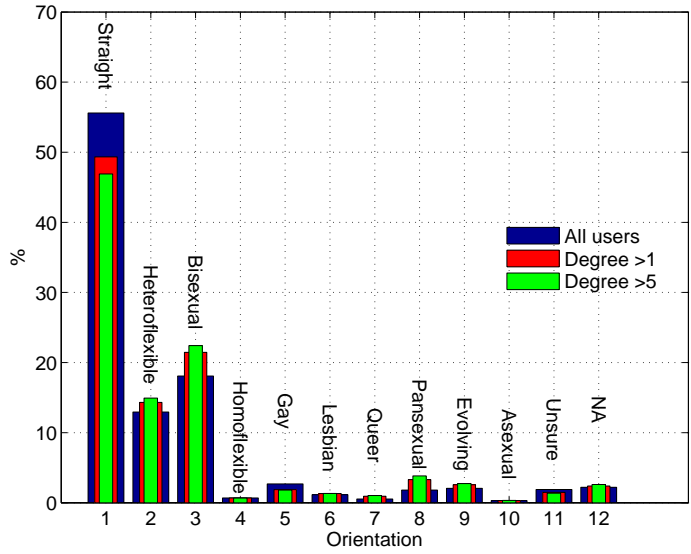} 
 \caption{Distribution of orientations for all users, users with $>1$ friends, and $>5$ friends.}\label{f:Orientation}
 \end{figure}

In this section we document the demographics of the fetish network such as gender, sexual preference, and connections. The identity acronyms are defined as follows: M = cis male; F = cis female; TV = transvestite; TS = transsexual, which can be further distinguished into male-to-female transsexuals (MtF or trans females) and female-to-male transsexuals (FtM or trans males); Kajira/Kajiru are slave girl/boy; I = intersex, B = butch, Fem = Femme. If not otherwise stated, Trans = trans females and TVs. GF = gender fluid and GQ = gender queer, referring to persons who do not identify as male or female or see themselves as having aspects of both genders. We first look at the gender demographics of the users as a whole. As mentioned previously, there are larger number of users with no friends than would otherwise be expected. Figure~\ref{f:gender_per} shows the distribution of user gender for all users. When the singletons have been removed, the gender distribution changes drastically; most of those with few or no friends are male (Figure~\ref{f:Orientation} shows that in addition they tend to be heterosexual males). When we have taken out those with fewer than 5 friends then the gender distribution is quite even with (cis) 54\% male, 40.5\% female  and other (non-cis) genders making up the remainder. Figure~\ref{f:GenderDrawing} diagrammatically is a graph indicative of the potential partners of different genders taking orientations into account. The graph is quite complicated with heterosexual relationships being reciprocal, gay relationships being homophilic (manifesting as graph loops), several uni-directional links (ex: a lesbian may consider a straight girl as a potential partner but this may not be reciprocated). In essence the sexual market which presents itself is neither bipartite nor undirected and so defies OSN analysis such as that in~\cite{Jerome}.

%
%
%


In Table~\ref{t:congruence} we examine the congruence of users with respect to gender and orientation. The network congruence is defined in~\cite{Singla} as the probability of one's friends having the same attributes or related attributes. That is, we wish to ask if people of a particular gender and orientation have preference for another gender. The results here show a strong preference in accordance with the graph shown in Figure~\ref{f:GenderDrawing}. For example, gay men have on average 32\% of their friends composed of gay men, far exceeding the population average of 1.5\%. A straight female will have 57\% of her friends as straight males, higher than the population average (39\%), slightly higher than the bisexual female average (53.4\%) and significantly higher than the gay female average (42\%). Overall the platonic relationships (in blue) are mostly lower than the population averages (exceptions are gay females who have a slightly higher than population average friendship with straight males; and gay trans to gay females). For straight females 73\% of their friends are straight/bisexual males. For straight males, 61.2\% of their friends are (straight/bisexual) females. This would strongly suggest a sexual market (for hetero- and bi-sexual people) as it implies not only a bias towards the opposite sex but also competition (see~\cite{Buss} for an excellent discussion). It implies that a male is less likely to be friends with the \emph{male friends of his female friends} than he would with a person from the population as a whole. That is, there would appear to be evidence of competition between males (and vice versa; also between females). This behaviour online  complements existing research in the literature that shows atypical sexual interests are more common in men than in women~\cite{dawson2014paraphilic}. 


\begin{table*}[ht]
 \resizebox{.8\columnwidth}{!}{
\begin{tabular}{|c|c|c|c|c|c|c|c|c|c|}
\hline
          &  M-S & M-Bi &  M-G &  F-S & F-Bi & F-G & Tr-S & Tr-Bi & Tr-G \\
      M-S & \textcolor{blue}{27.1} &  \textcolor{blue}{7.1} &  \textcolor{blue}{0.7} &\textbf{17.1} & 44.1 & \textcolor{blue}{1.5} &     \textcolor{blue}{0.3} &  \textcolor{blue}{1.9} &\textcolor{blue}{0.3} \\
     M-Bi & 26.8 & 12.9 &  3.0 & 12.7 & 36.4 & 1.4 &     0.6 &      5.6 &     0.6 \\
      M-G & 23.8 & 27.6 & \textbf{32.9} &  \textcolor{blue}{2.7} &  \textcolor{blue}{8.4} & \textcolor{blue}{0.8} &     0.2 &      3.1 &     0.5 \\
      F-S & \textbf{57.0} & 16.3 &  \textcolor{blue}{0.4} &  \textcolor{blue}{6.5} & \textcolor{blue}{14.9} & \textcolor{blue}{0.9} &    \textcolor{blue}{0.6} & \textcolor{blue}{3.2} &\textcolor{blue}{0.3} \\
     F-Bi & 53.4 & 16.8 &  0.4 &  5.4 & 18.7 & 1.4 &     0.5 &      3.1 &     0.3 \\
      F-G & \textcolor{blue}{42.9} & \textcolor{blue}{12.9} &  \textcolor{blue}{0.8} &  6.2 & 26.8 & \textbf{6.1} &     0.5 &      3.0 &     0.8 \\
  Tr-S & \textcolor{blue}{25.6} & \textcolor{blue}{12.2} &  \textcolor{blue}{0.6} &  9.4 & 22.8 & 1.1 &     3.5 &     22.9 &     1.8 \\
 Tr-Bi & 23.9 & 15.7 &  1.0 &  7.0 & 18.7 & 0.9 &     3.0 &     27.6 &     2.3 \\
  Tr-G & 29.9 & 16.1 &  1.5 &  \textcolor{blue}{5.8} & \textcolor{blue}{17.8} & \textcolor{blue}{2.3} &     2.2 &     21.4 &    \textbf{3.0} \\ \hline
      All & 39.1 & 13.2 &  1.5 & 10.3 & 28.8 & 1.4 &     0.6 &      4.7 &     0.5 \\ \hline    \end{tabular}\caption{Congruency of gender and orientation: $\{Male,Female,Trans\} \times \{Straight,Bisexual,Gay\}$. Potential partners in black, platonic in blue, and conventional partners in \textbf{bold}.}\label{t:congruence}
 }
 \end{table*}
 
We compared our results with that of Pokec, a large European OSN of over 1.6 million subscribers with gender specifications~\cite{Takac2012}. In Pokec, male members are 49\% and 51\% likely to connect to males and females respectively, while these figures are 55\% and 45\% for females connecting to females and males respectively. This is a rather balanced ratio and in a rather significant contrast with the fetish network's data which has a strong bias towards the opposite sex, further supporting the sexual market social network hypotheses. It is worth noting that, although men are more active users of cybersex channels, significantly more women than men state that their online sexual activities had led to real-life sexual encounters~\cite{Schneider2000}. 

For the TV, MtF, FtM, and TG users there appears to be a strong preference towards friends of the same gender. For example, a TV will tend to have 29.5\% friends, far above the population average of 4.7\%. However, it is interesting to note that while there is a strong bias towards people of the same gender the majority of friends still come from other genders; there is no evidence to support the idea of closed minority gender communities. 

Figure~\ref{f:Orientation} shows the distribution of sexual orientations of users. Of the users, 45\% describe themselves as heterosexual while less than 5\% describe themselves as exclusively gay or lesbian. Large survey-based studies show that BDSM activities are more common among non-heterosexual individuals (gay, lesbian or bisexual)~\cite{Richters2008}.

 \begin{figure}[t!]
 \centering
 \includegraphics[width=0.75\textwidth]{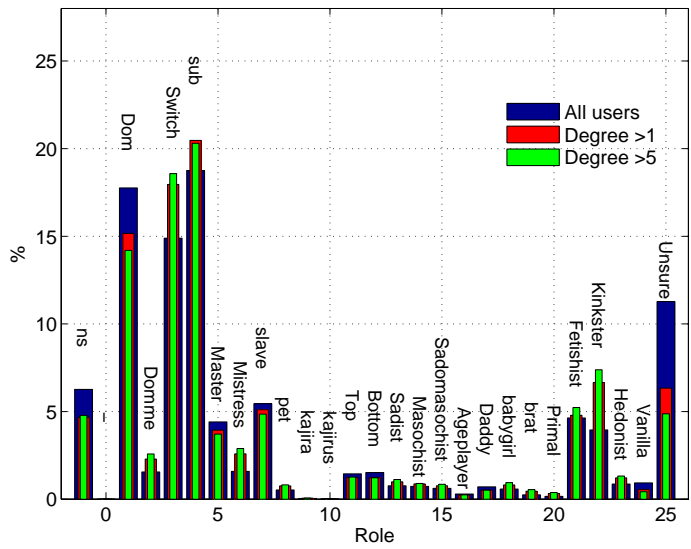} 
 \caption{Distribution of roles for all users, users with $>1$ friends, and $>5$ friends.}\label{f:Role}
 \end{figure}
 
%
%


Figure~\ref{f:Role} shows the distribution of user roles. Some of these roles are quite similar and interestingly the dominant roles (Dom, Domme, Mistress, Master) take up 23.39\% of the roles while the submissive roles (Sub, Slave) take up 25.15\% of the users, a remarkable balance. 

\begin{figure}[t!]
 \centering
 \includegraphics[scale=0.45]{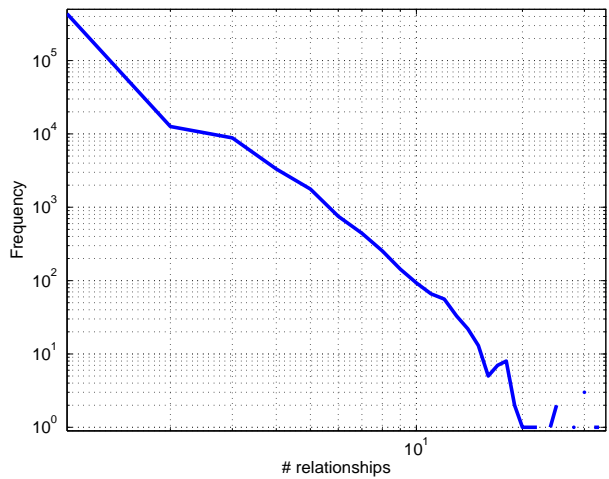}
 \caption{Distribution of number of relationships.}\label{f:rel_deg_dist}
 \end{figure}

Figure~\ref{f:rel_deg_dist} shows the distribution of the number of relationships that individuals are in. The largest number of relationships amongst all users is involved in is 30 and the distribution of the other relationships is power law. We highlight that these are declared relationships and visible to all members.

 \section{Network analysis.}
 \label{sec:net_anal}
\begin{figure}
\begin{floatrow}
\ffigbox{  \includegraphics[width=0.48\textwidth]{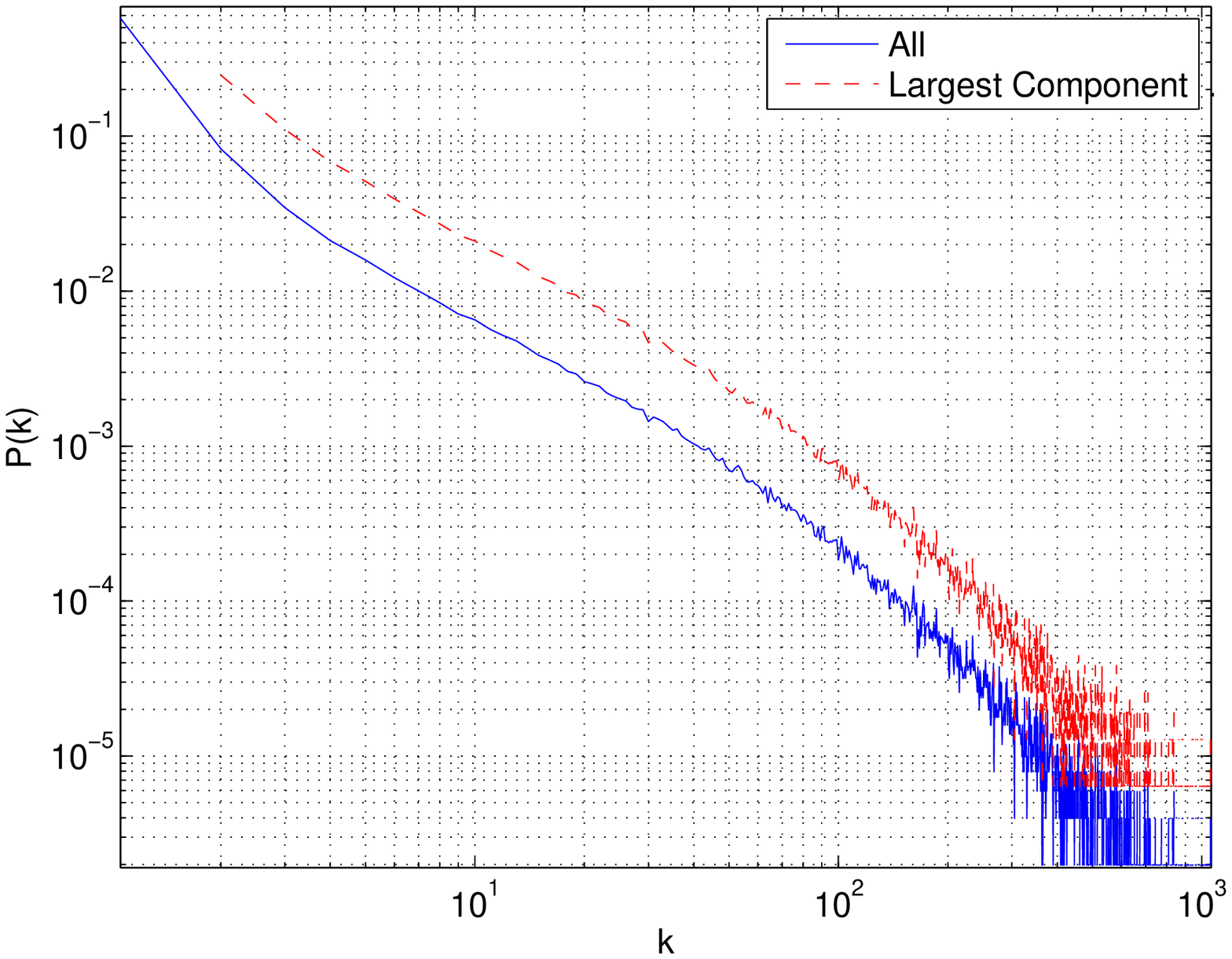} }
{%
\caption{Degree distribution of the network and LCC.}\label{f:deg_dist}
  
}
\ffigbox{  \includegraphics[width=0.48\textwidth]{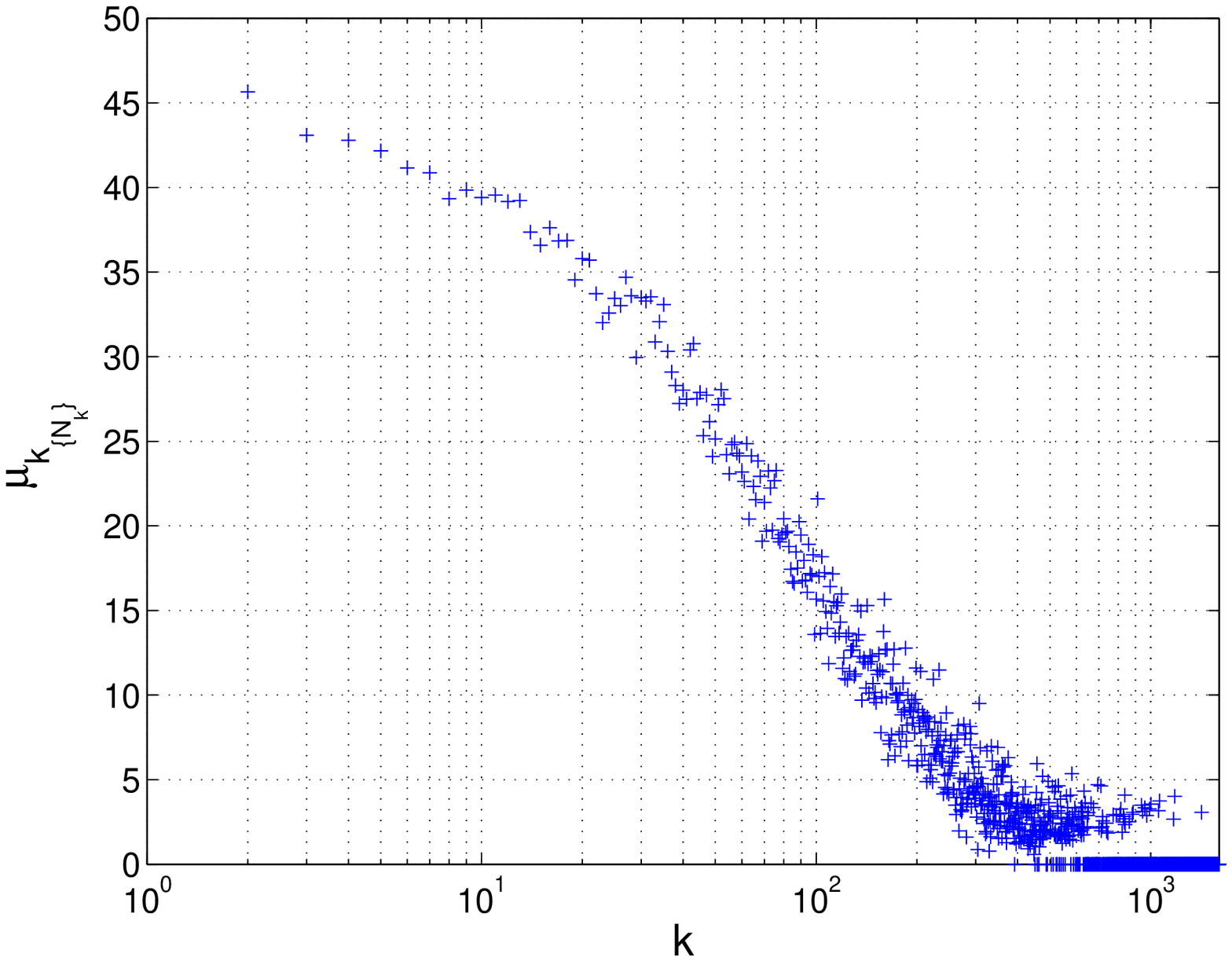} }
{%
\caption{Semi-log plot of the average degree of the friends of users with degree of $k$. }\label{f:knn}
  
}

\end{floatrow}
\end{figure}

For comparison we examined the fetish network structure with those of standard OSNs (YouTube, Flickr, LiveJournal and Orkut) following the analysis, and using results, of~\cite{Mislove}. 
We then look into more complex measures such as the average path length, Joint Degree Distribution (JDD, a measure of connectivity of one's neighbours),  clustering coefficient (measure of density of triangular ties between adjacent nodes), and assortativity, which indicate the relations between the nodes on a local basis. We also explore the hierarchical structure of the network using k-cores and Kernel density estimation.\footnote{A complete explanation of the theoretical definitions and implications of these measures is available in~\cite{fay2010weighted} and~\cite{haddadi2008tuning}.} The degree distribution is shown in Figure~\ref{f:deg_dist}, and is unremarkable except that there is a larger than expected number of users with low degree. These are removed when we examine the main component of the graph (as previously mentioned these users would appear to be \emph{lurkers}; mostly heterosexual males who do not participate in the social network). Figure~\ref{f:knn} shows the distribution of the degrees of friends of degree $k$. Again this is unremarkable and similar may be found in~\cite{Mislove}. Finally, Table~\ref{t:musires} gives a summary of common network measures.\footnote{We assume that the reader is familiar with standard network measures (a good overview may be found in \cite{Kleinberg_book,fay2010weighted,haddadi2008tuning})} The main conclusion is that FetLife has a very similar structure to most OSN's.   

\begin{table*}[ht] 
{\scriptsize 
    \centering
    \resizebox{\columnwidth}{!}{
    \begin{tabular}{|l|c|c|c|c|c|c|}
      \hline
     &     $\alpha$   &     avg path length   &     Radius|Diameter  &  assortativity &   scale free metric &    avg clustering coefficient  \\
 Fetish & 2.98 & 4.05 &  7|11 & -0.01 & 0.0031 &  0.15 \\
   Flickr & 1.74 & 5.67 &  13|27 & 0.202 & 0.49 & 0.313  \\
   Livejournal & 1.59 & 5.88 &  12|20 & 0.179 & 0.34 & 0.330  \\
  Orkut & 1.50 & 4.25 & 6|9 & 0.072 & 0.36 & 0.171  \\
 Youtube & 1.63 & 5.10 & 13|21 & -0.033 & 0.19 & 0.136  \\
 Web & 2.57 & 16.12 & 475|905 & -0.067 & - & 0.081  \\
  \hline
\end{tabular}
} 
    \caption{\footnotesize Network Measures from the fetish and ordinary OSNs.\label{t:musires}}}
\end{table*}

\begin{figure}
 \begin{floatrow}
 \ffigbox{  \includegraphics[width=0.45\textwidth]{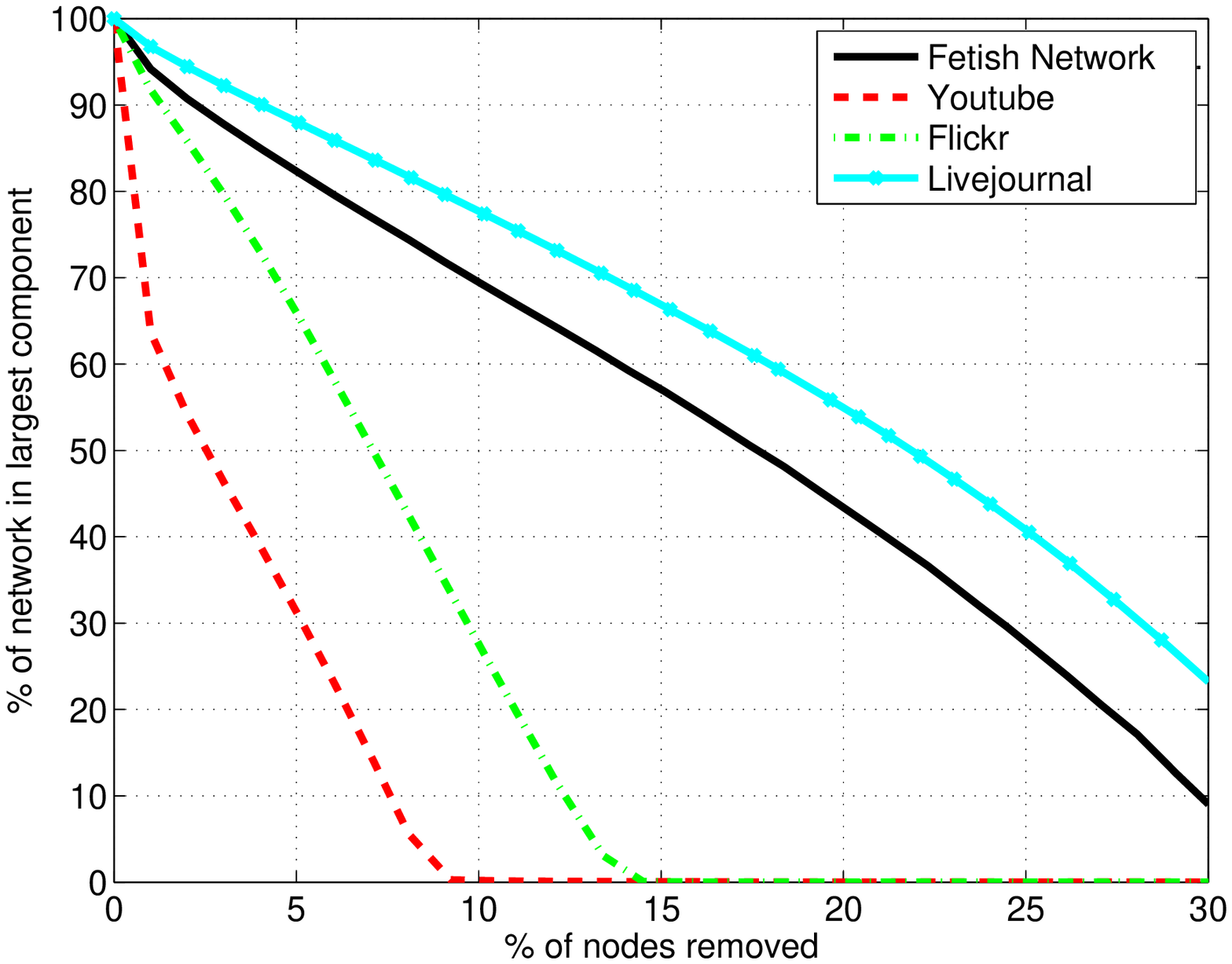} }
 {%
 \caption{Percentage of main component remaining after removal of the highest degree users. }\label{f:kcore}
   
 }
 \ffigbox{  \includegraphics[width=0.45\textwidth]{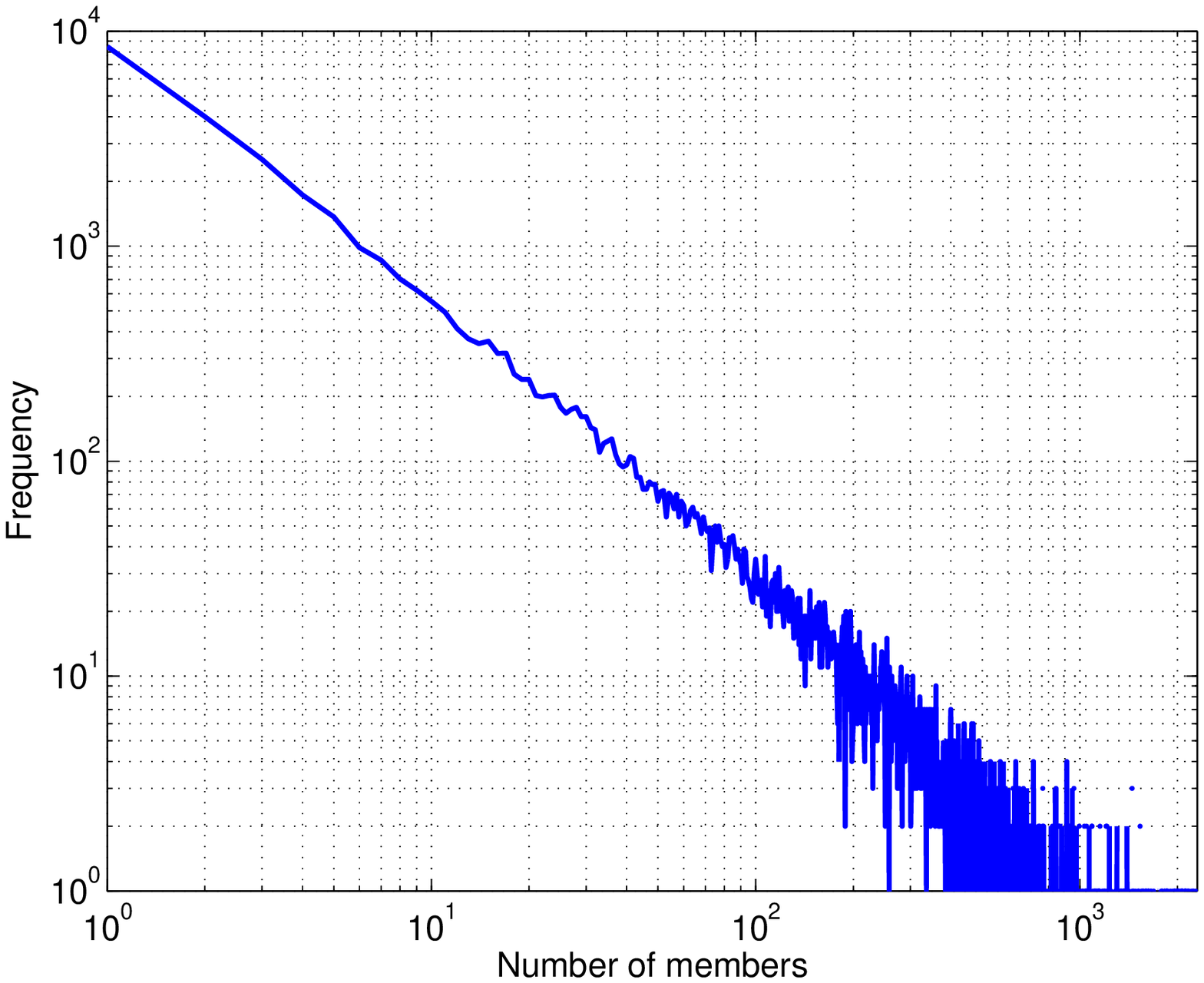} }
 {%
 \caption{loglog plot of Group sizes with number of members. }\label{f:group_power_law}
   
 }
 
 \end{floatrow}
 \end{figure}

Figure~\ref{f:kcore} shows the k-core of removal rate and that the network is highly resilient to removal of high degree nodes. In fact we could remove the top 10\% of the nodes and only lose 30\% off the largest connected component. This indicates that the network consists of lots of small connections between people ignoring the core. The large number of small groups and local clusters, as opposed to large inter-mixed nodes, is the main reason behind this effect, which has also recently been observed in the Internet topology~\cite{haddadi2010mixing}. In FetLife, the events and connections are centred around local events, meetings, and workshops. Although a direct search function is not available, many users of the website use the network as a portal to bootstrap their fetish sex life. Hence the \emph{global} connectivity is not as important as traditional OSNs such as Twitter and Facebook, and far from \emph{content}-centric OSNs such as Flickr and YouTube.


 Our analysis consists of two main parts. In this section we compare the fetish network with topologies of datasets from mainstream  to ascertain if this network is a social network form structural point of view. 
 
 \subsection*{Network structure}
 \label{subsec:net_struct}


 Figure~\ref{f:deg_dist} shows the degree distribution for the entire network and the largest connected component. There is a larger number of users with degree $<5$ than would be expected from extrapolating the distribution from the right (see Section~\ref{subsec:demo_anal} for details). Once we examine the largest connected component the distribution takes on a more standard shape in fitting with the standard OSN distributions reported by Mislove~\etal~\cite{Mislove}. Each user has on average 24 friends in the largest connected component  (8 if we include all users) again in rough agreement (12 Flickr, 17 Livejournal and 106 for Orkut) with the OSNs reported in~\cite{Mislove}.\footnote{Aside: the average number of friends can have a large variance and can increase to over 100 if we exclude people with only a single friend, this is a symptom of the fact that the definition of an average for a power-law distribution is largely uninformative~\cite{li_towards}}. One explanation for the smaller number of friends and flatter networks in Fetlife versus otherOSNs is that those other networks draw on relationships outside the network, e.g., old school friends, co-workers, and extended family. Here, some individuals have outside friendships that translate to Fetlife, but many are likely to be at least partially closeted, and thus have fewer social connections to draw on. Instead, friendships are built online. 
 
 We now consider the unbiased power-law exponent~\cite{clauset07-1} estimated from the data in figure~\ref{f:deg_dist}. The isolated members (those with 0 friends) in the main network are predominantly male (see Section~\ref{subsec:demo_anal}). The power law coefficient, $\alpha$, is 2.98 (3 including singletons). This number is comparable to the web out-degree (2.67) but higher than that of the standard OSNs (see Table~\ref{t:musires}). This indicates a relatively larger number of high degree users in the fetish network than in the standard OSN. The average path length (4.05), radius (7.00) and diameter (11) are however closest to the Orkut graph and the graph is much more densely connected than the web. Hence these basic measures are not enough for understanding the nature of this network. 
 The graph has a stronger power-law coefficient than most conventional social networks.

 \begin{figure}
 \begin{floatrow}
 \ffigbox{  \includegraphics[width=0.45\textwidth]{images/knn.eps} }
 {%
 \caption{Semi-log plot of the average degree of the friends of users with degree of $k$. }\label{f:knn}
   
 }
 \ffigbox{  \includegraphics[width=0.45\textwidth]{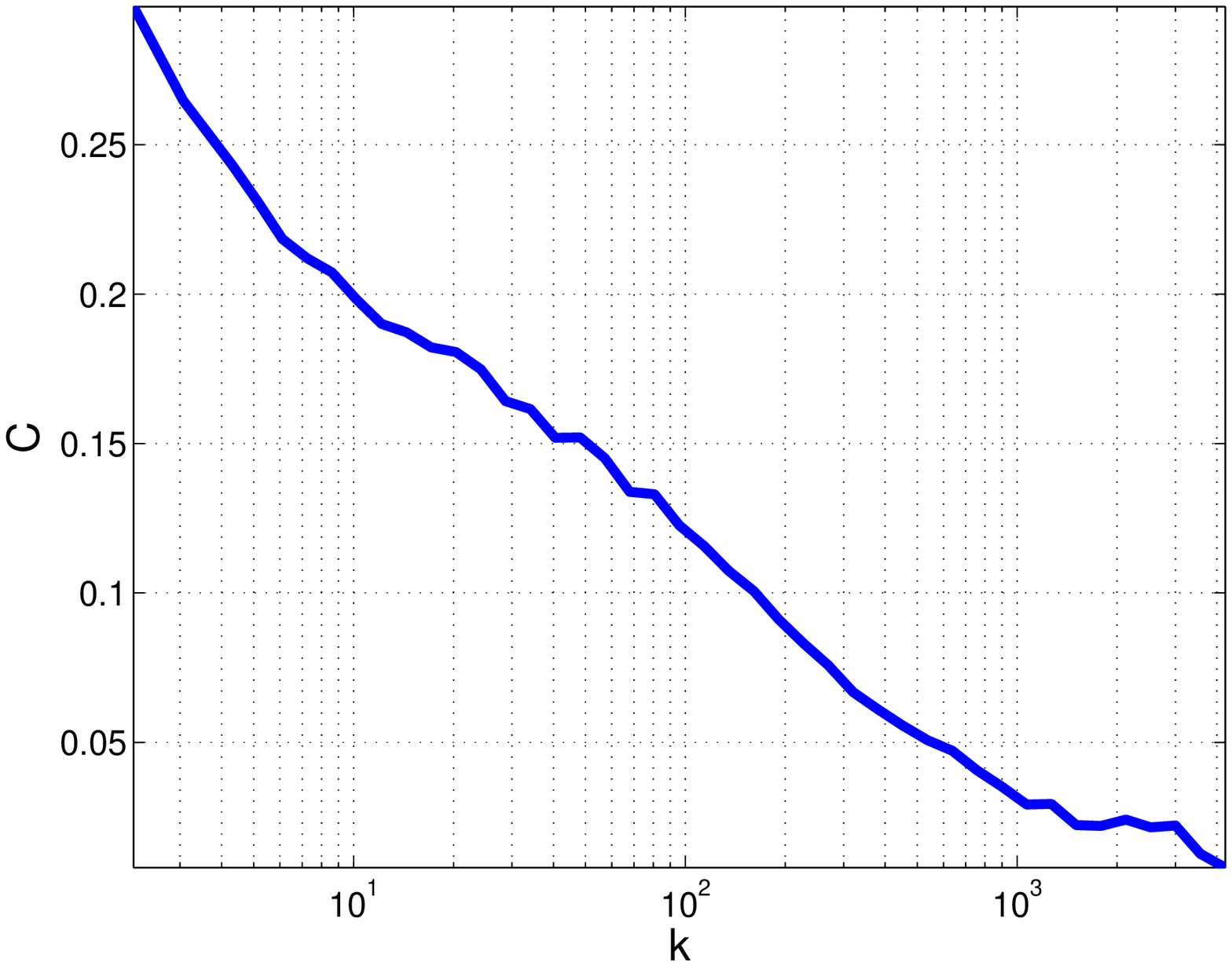} }
 {%
 \caption{Clustering coefficient distribution as a function of user degree, $k$. }\label{f:ccfs}
   
 }
 
 \end{floatrow}
 \end{figure}


 Figure~\ref{f:knn} shows the average degree of the friends of users with degree of $k$. The decrease in JDD indicates that individuals with few connections tend to have friends with high degree, rather than connecting to low degree users. This is similar to the YouTube or the web network where many users connect to popular nodes, and dissimilar to traditional OSNs where ordinary individuals connect to each other. This trend is confirmed by the slightly negative assortativity (-0.01, i.e., dis-assortativity) of the the network. On close inspection of the popular accounts, this effect seems to be due to the different motivation of users with a high degree; based on a manual random sample we found these users to be fetish fashion businesses, photographers or nightclubs and it may be that their aims are work related. The Scale-free metric of the JDD~\cite{li_towards} is also 0.0031, much lower than Flickr (0.49), LiveJournal (0.34), Orkut (0.36), and YouTube (0.19); this indicates that there is a higher hub and spikes structure in FetLife, where high degree nodes tend to connect to low degree nodes, acting as mediators and organisers of events.


 Despite the high number of users and links, the average clustering coefficient of the network is 0.1544, which is lower than all OSNs reported in~\cite{Mislove}, this is mainly due to the \textit{introductory} nature of the network and its groups: the main objective of the users in the network is to find new partners and arrange events in real life, hence the group has a strong hierarchy. Figure~\ref{f:ccfs} displays the distribution of clustering coefficients for the groups after removing singletons (26k nodes). The distribution is similar to a social network where individuals with lower connectivity are more likely to be in tight groups with their connections, rather than being part of larger groups. 

\section{Homophilic Community Detection}
\label{sec:community}
From the analysis above (Table~\ref{t:congruence} in particular) we see a network where there are connections for many reasons. Some connections are created for sexual attraction, others are purely social. Within the sexual attractions there is homophilic and heterophilic factors and in addition there are heterophilic sexual connections to do with a persons role (a dominant person would in particular like a submissive person).
It is possible to detect and separate homophilic communities from heterophilic communities to gain insights into the nature of homophilic relations in the network while factoring out heterophilic relations.
Homophilic community detection is a complicated task requiring not just knowledge of the links in the network but also the attributes associated with those links. A recent paper by Yang et. al.~\cite{cesna} proposed the CESNA model (Community Detection in Networks with Node Attributes). This model is generative and based on the assumption that a link is created between two users if they share membership of a particular community. Users within a community share similar attributes.
Therefore, the model is able to extract homophilic communities from the link network.
Vertices may be members of several \emph{independent} communities such that the probability of creating an edge is 1 minus the probability that no edge is created in any of their common communities: 
\begin{equation}
P_{u \rightarrow v} =  1 - \prod_{c \in C} \exp({ -F_{uc} \cdot F_{vc}})
\end{equation}\label{e:1}
where $F_{uc}$ is the potential of vertex $u$ to community $c$ 
and $C$ is the set of all communities. In addition, it assumed that the attributes of a vertex are also generated from the communities they are members of and so the graph and the attributes are generated \emph{jointly} by some underlying unknown community structure. Specifically the attributes are assumed to be binary (present or not present) and are generated according to a Bernoulli process: 
\begin{equation}
X_{uk} \sim \mathcal{B} \Big( Q_k \Big) 
\end{equation}
where $Q_k = 1 / \left( {1+ \prod_{c \in C} \exp({{- W_{kc}F_{uc}  } })}\right)$, $W_{kc}$ is a weight matrix $\in \mathbb{R}^{N\times|C|}$,\footnote{There is also a bias term $W_0$ which has an important role. We set this to -10; otherwise if someone has a community affiliation of zero, $F_u=0$, $Q_k$ has probability $\frac{1}{2}$.}  which defines the strength of connection between the $N$ attributes and the $|C|$ communities. $W_{kc}$ is central to the model and is a set of logistic model parameters which -- together with the number of communities, $|C|$ -- forms the set of unknown parameters for the model. Parameter estimation is achieved by maximising the likelihood of the observed graph (i.e. the observed connections) and the observed attribute values given the membership potentials and weight matrix. As the edges and attributes are conditionally independent given $W$, the log likelihood may be expressed as a summation of three different events: 

\begin{equation}
 log P(G,X|F,W) = \sum_{u,v\in E} \log(1-e^{-F_u F_v^T}) - \sum_{u,v\notin E} F_u F_v^T \qquad  \qquad  \\ 
 \end{equation} 
 \begin{equation*}
 \qquad \qquad  \qquad  \qquad  \qquad + \sum_{u,k} \left[ X_{uk} \log(Q_{uk}) + (1-X_{uk}) \log(1-Q_{uk}) \right]
 \end{equation*} 
where the first term on the right hand side is the probability of observing the edges in the network, the second term is the probability of observing the non-existent edges in the network, and the third term are the probabilities of observing the attributes under the model. 
An inference algorithm is given in~\cite{cesna}.
The data used in the community detection for this network consists of the main component of the network together with the attributes \{\textit{Male, Female, Trans, GQ}\} together with orientations \{\textit{Straight, Bisexual, Gay}\} and roles \{\textit{submissive, dominant, switch}\} for a total of 10 binary attributes. We found that, due to large imbalance in the size of communities, we needed to generate a large number of communities before observing the niche communities (e.g. trans and gay). Generating communities varying $|C|$ from 1 to 50, we observed the detected communities persist as $|C|$ grows or split into two communities (i.e as $|C|$ increases we uncover a natural hierarchy). Table~\ref{t:weights} shows the attribute probabilities for each community, specifically: $Q_k|_{F_u=10}$. For analysis we have grouped these communities into \emph{Super-Communities} (SC's) based on common attributes. 

The first five SC's are for a single gender alone (GQ, cis male, and cis female; SC3 and SC4, SC5). SC2 consists of only bi or gay males, mostly gay males, and the absence of any straight male (alone) group is very apparent. SC3 consists of straight and bi (cis) females, SC4; all cis females, and SC5; only gay females (i.e. lesbian). SC6 consists mainly of cis females (GQ account for only 1\% of the population.). There is therefore very strong evidence of many communities of (i.e. complex) female to female interaction that is largely platonic. In SC8 the transgender community appears clearly. SC10 is the only community to contain straight (cis) males and straight females together and accounts for only 3.6\% of those classified. SC11 and SC12 shows interaction between cis females and trans members which accounts for at least 8\% of those classified. The above shows complex interactions between the members, some are expected (trans and gay specific communities) while the absence of straight males from all but a small community is stark. 

 \begin{table*}[ht!]
 \scriptsize
 \resizebox{.8\columnwidth}{!}{
\begin{tabular}{|c||c|c|c|c||c|c|c||c|c|c||c|}
\hline
Attribute & M & F &  Tr &  GQ & Str8 & Bi & Gay & Dom & Sub & Switch & \% \\
\hline\hline
\multicolumn{11}{ |c  }{Supercommunity SC1} & 2.90 \\ \hline
C32 & 0.0 & 0.0 & 0.0 & 1.0 & 1.0 & 1.0 & 0.0 & 0.0 & 1.0 & 1.0 & 2.90 \\ \hline\hline
\multicolumn{11}{ |c  }{Supercommunity SC2} & 2.70 \\ \hline
C38 & 1.0 & 0.0 & 0.0 & 1.0 & 0.0 & 0.0 & 1.0 & 0.0 & 1.0 & 0.0 & 2.69 \\
C49 & 1.0 & 0.0 & 0.0 & 0.0 & 0.0 & 1.0 & 0.0 & 0.0 & 1.0 & 0.0 & 0.01 \\ \hline\hline
\multicolumn{11}{ |c  }{Supercommunity SC3} &  21.00 \\ \hline 
C2 & 0.0 & 1.0  & 0.0 & 0.0 & 1.0 & 1.0 & 0.0 & 0.0 & 0.0 & 0.0 & 4.50 \\
C14 & 0.0 & 1.0 & 0.0 & 0.0 & 1.0 & 1.0 & 0.0 & 0.0 & 0.0 & 0.0 & 3.90 \\
C20 & 0.0 & 1.0 & 0.0 & 0.0 & 1.0 & 1.0 & 0.0 & 1.0 & 0.0 & 1.0 & 3.91 \\
C30 & 0.0 & 1.0 & 0.0 & 0.0 & 1.0 & 1.0 & 0.0 & 0.0 & 0.0 & 1.0 & 2.16 \\
C34 & 0.0 & 1.0 & 0.0 & 0.0 & 1.0 & 1.0 & 0.0 & 1.0 & 1.0 & 1.0 & 2.69 \\
C43 & 0.0 & 1.0 & 0.0 & 0.0 & 1.0 & 1.0 & 0.0 & 0.0 & 0.0 & 1.0 & 2.42 \\
C45 & 0.0 & 1.0 & 0.0 & 0.0 & 1.0 & 1.0 & 0.0 & 0.0 & 0.0 & 1.0 & 2.11 \\
C48 & 0.0 & 1.0 & 0.0 & 0.0 & 1.0 & 1.0 & 0.0 & 0.0 & 0.0 & 0.0 & 1.49 \\ \hline\hline
\multicolumn{11}{|c  }{Supercommunity SC4 } &  16.42 \\ \hline
C1 & 0.0 & 1.0  & 0.0 & 0.0 & 1.0 & 1.0 & 1.0 & 0.0 & 0.0 & 1.0 & 6.18 \\
C15 & 0.0 & 1.0 & 0.0 & 0.0 & 1.0 & 1.0 & 1.0 & 0.0 & 0.0 & 0.0 & 4.40 \\
C7 & 0.0 & 1.0  & 0.0 & 0.0 & 1.0 & 1.0 & 1.0 & 0.0 & 0.0 & 1.0 & 4.33 \\
C26 & 0.0 & 1.0 & 0.0 & 0.0 & 1.0 & 1.0 & 1.0 & 0.0 & 0.0 & 1.0 & 2.96 \\ \hline\hline
\multicolumn{11}{ |c  }{Supercommunity SC5} &  6.36 \\ \hline
C16 & 0.0 & 1.0 & 0.0 & 0.0 & 0.0 & 0.0 & 1.0 & 1.0 & 1.0 & 1.0 & 3.29 \\
C8 & 0.0 & 1.0  & 0.0 & 0.0 & 0.0 & 0.0 & 1.0 & 1.0 & 1.0 & 1.0 & 4.21 \\\hline\hline
\multicolumn{11}{ |c  }{Supercommunity SC6} &  48.10 \\ \hline
C27 & 0.0 & 1.0 & 0.0 & 1.0 & 1.0 & 0.0 & 0.0 & 0.0 & 0.0 & 1.0 & 3.66 \\
C10 & 0.0 & 1.0 & 0.0 & 1.0 & 1.0 & 0.0 & 0.0 & 0.0 & 0.0 & 0.0 & 4.63 \\
C11 & 0.0 & 1.0 & 0.0 & 1.0 & 1.0 & 0.0 & 0.0 & 0.0 & 0.0 & 1.0 & 3.83 \\
C12 & 0.0 & 1.0 & 0.0 & 1.0 & 1.0 & 0.0 & 0.0 & 0.0 & 0.0 & 0.0 & 4.79 \\
C17 & 0.0 & 1.0 & 0.0 & 1.0 & 1.0 & 0.0 & 0.0 & 0.0 & 0.0 & 1.0 & 3.52 \\
C18 & 0.0 & 1.0 & 0.0 & 1.0 & 1.0 & 0.0 & 0.0 & 0.0 & 0.0 & 0.0 & 4.86 \\
C19 & 0.0 & 1.0 & 0.0 & 1.0 & 1.0 & 0.0 & 0.0 & 0.0 & 0.0 & 1.0 & 3.64 \\
C21 & 0.0 & 1.0 & 0.0 & 1.0 & 1.0 & 0.0 & 0.0 & 1.0 & 0.0 & 1.0 & 3.64 \\
C22 & 0.0 & 1.0 & 0.0 & 1.0 & 1.0 & 0.0 & 0.0 & 0.0 & 0.0 & 1.0 & 2.85 \\
C28 & 0.0 & 1.0 & 0.0 & 1.0 & 1.0 & 0.0 & 0.0 & 0.0 & 0.0 & 1.0 & 2.97 \\
C29 & 0.0 & 1.0 & 0.0 & 1.0 & 1.0 & 0.0 & 0.0 & 0.0 & 1.0 & 1.0 & 2.73 \\
C33 & 0.0 & 1.0 & 0.0 & 1.0 & 1.0 & 0.0 & 0.0 & 0.0 & 0.0 & 1.0 & 3.43 \\
C35 & 0.0 & 1.0 & 0.0 & 1.0 & 1.0 & 0.0 & 0.0 & 1.0 & 0.0 & 1.0 & 2.62 \\
C36 & 0.0 & 1.0 & 0.0 & 1.0 & 1.0 & 0.0 & 0.0 & 0.0 & 0.0 & 1.0 & 2.71 \\
C37 & 0.0 & 1.0 & 0.0 & 1.0 & 1.0 & 0.0 & 0.0 & 0.0 & 1.0 & 1.0 & 2.46 \\
C39 & 0.0 & 1.0 & 0.0 & 1.0 & 1.0 & 0.0 & 0.0 & 0.0 & 0.0 & 1.0 & 2.47 \\
C41 & 0.0 & 1.0 & 0.0 & 1.0 & 1.0 & 0.0 & 0.0 & 0.0 & 0.0 & 0.0 & 2.74 \\
C44 & 0.0 & 1.0 & 0.0 & 1.0 & 1.0 & 0.0 & 0.0 & 0.0 & 1.0 & 1.0 & 2.53 \\
C47 & 0.0 & 1.0 & 0.0 & 1.0 & 1.0 & 0.0 & 0.0 & 0.0 & 1.0 & 1.0 & 1.77 \\ \hline\hline
\multicolumn{11}{ |c  }{Supercommunity SC7} &  10.16 \\ \hline
C3  & 0.0 & 1.0 & 0.0 & 1.0 & 1.0 & 0.0 & 1.0 & 0.0 & 0.0 & 0.0 & 5.10 \\
C13 & 0.0 & 1.0 & 0.0 & 1.0 & 1.0 & 1.0 & 0.0 & 0.0 & 0.0 & 1.0 & 3.41 \\
C40 & 0.0 & 1.0 & 0.0 & 1.0 & 1.0 & 1.0 & 0.0 & 0.0 & 0.0 & 1.0 & 1.88 \\ \hline\hline
\multicolumn{11}{ |c  }{Supercommunity SC8} &  3.61 \\ \hline
C23 & 0.0 & 0.0 & 1.0 & 1.0 & 1.0 & 1.0 & 0.0 & 0.0 & 0.0 & 1.0 & 3.61 \\ \hline\hline
\multicolumn{11}{ |c  }{Supercommunity SC9} &  11.24 \\ \hline
C31 & 1.0 & 1.0 & 0.0 & 0.0 & 0.0 & 0.0 & 1.0 & 1.0 & 1.0 & 0.0 & 3.59 \\
C5 & 1.0 & 1.0 & 0.0 & 0.0 & 0.0 & 1.0 & 0.0 & 0.0 & 0.0 & 1.0 & 3.61 \\
C42 & 1.0 & 1.0 & 0.0 & 0.0 & 0.0 & 0.0 & 1.0 & 0.0 & 1.0 & 0.0 & 2.47 \\
C46 & 1.0 & 1.0 & 0.0 & 0.0 & 0.0 & 1.0 & 0.0 & 0.0 & 1.0 & 1.0 & 2.43 \\ \hline\hline
\multicolumn{11}{ |c  }{Supercommunity SC10} &  3.59 \\ \hline
C24 & 1.0 & 1.0 & 0.0 & 1.0 & 1.0 & 1.0 & 0.0 & 0.0 & 0.0 & 1.0 & 3.58 \\
C50 & 1.0 & 1.0 & 0.0 & 1.0 & 0.0 & 0.0 & 0.0 & 0.0 & 0.0 & 0.0 & 0.01 \\ \hline\hline
\multicolumn{11}{ |c  }{Supercommunity SC11} & 8.02 \\ \hline
C6 & 0.0 & 1.0 & 1.0 & 0.0 & 1.0 & 1.0 & 1.0 & 0.0 & 0.0 & 0.0 & 4.72 \\
C25 & 0.0 & 1.0 & 1.0 & 0.0 & 1.0 & 1.0 & 1.0 & 0.0 & 0.0 & 0.0 & 3.34 \\\hline\hline
\multicolumn{11}{ |c  }{Supercommunity SC12} & 8.25 \\ \hline
C4 & 0.0 & 1.0 & 1.0 & 1.0 & 1.0 & 1.0 & 0.0 & 0.0 & 0.0 & 1.0 & 3.48 \\
C9 & 0.0 & 1.0 & 1.0 & 1.0 & 1.0 & 0.0 & 0.0 & 0.0 & 0.0 & 0.0 & 5.04 \\ \hline
\end{tabular}\caption{$Q_K$ for 50 communities and the percentage of the population classed in the community using 10 attributes.}\label{t:weights}
 }
 \end{table*}

\section{Conclusion}
\label{sec:conclusion}

In  this paper, we conducted the first in-depth study of a large fetish community, exploring not only the attributes of users but also analysing the rich structures behind the social network of community members. 

The diversity of online sexual contacts is of growing importance~\cite{user_demographics} and several studies have explored how the diversity of sexual contacts is affecting sexuality~\cite{Döring20091089}.
The fetish network examined in this paper has many of the functionalities of mainstream OSNs: there are friendship links, relationships, interests, groups and events. The extracted user network is a valuable source of information as a fetish community is neither a dating website nor a standard OSN. Rather, it is an OSN where the sexual market aspects of the network have been amplified. The picture that emerges is one of complex hetero- and homo-philic interacting communities and in addition, we observed friendships which are purely platonic. 
We successfully extracted and analysed homophilic relations and communities from the network employing the CESNA community detection algorithm, paving the way for further studies on homophilic and heterophilic communities. The dearth of straight males in the homophilic communities perhaps makes sense; straight women in particular are on the site for platonic social reasons more than other groups. It would also makes sense that straight males are less so, i.e., they are more interested in potential sexual connections. Note this does not mean that \emph{all} straight men are not interested in platonic relationships; rather every straight man interested in \emph{only} sexual connections is a counter-example to the others and there is no extra information (one could imagine for example "\emph{platonic} straight male" ) to discriminate the two groups. 

In future work we will further investigate and stochastically model the complex community structures behind the social network, including additional profile information such as freely chosen tags by the users.



%


\bibliographystyle{abbrv}
\bibliography{topology}   

\end{thebibliography}

\end{document}